\documentclass[a4paper,11pt]{article}
\pdfoutput=1 

\usepackage{jcappub} 

\usepackage[T1]{fontenc} 
\usepackage[utf8]{inputenc}

\usepackage{graphicx}
\usepackage{booktabs}
\usepackage{amsmath}
\usepackage{amssymb}
\usepackage{amsthm}
\usepackage{hyperref}
\usepackage[caption=false]{subfig}
\usepackage{color,soul}
\usepackage{xcolor}
\usepackage{multirow}
\usepackage{float}

\usepackage{cancel}

\newcommand{\nver}{\hat{\mathbf{n}}}

\title{\boldmath Needlet estimation of cross-correlation between CMB lensing maps and LSS}

\author[a,e,d,1]{Federico Bianchini,\note{Corresponding author.}}
\author[b,c]{Alessandro Renzi,}
\author[b,c]{Domenico Marinucci}

\affiliation[a]{Astrophysics Sector, SISSA, Via Bonomea 265, I-34136 Trieste, Italia}
\affiliation[b]{Dipartimento di Matematica, Universit\'a di Roma Tor Vergata, Via della Ricerca 
  Scientifica 1, 00133 Roma, Italia}
\affiliation[c]{INFN, Sezione di Roma 2, Universit\'a di Roma Tor Vergata, Via della Ricerca Scientifica 
  1, 00133 Roma, Italia}
\affiliation[d]{INFN - Sezione di Trieste, Via Valerio 2, I-34127 Trieste, Italy}
\affiliation[e]{INAF - Osservatorio Astronomico di Trieste, via Tiepolo 11, 34131, Trieste, Italy}

\emailAdd{fbianchini@sissa.it}
\emailAdd{renzi@mat.uniroma2.it}
\emailAdd{marinucc@mat.uniroma2.it}

\abstract{In this paper we develop a novel needlet-based estimator to investigate the cross-correlation 
between cosmic microwave background (CMB) lensing maps and large-scale structure (LSS) data. We 
compare this estimator with its harmonic counterpart and, in particular, we analyze the bias effects of 
different forms of masking. In order to address this bias, we also implement a MASTER-like technique in 
the needlet case. The resulting estimator turns out to have an extremely good signal-to-noise 
performance. Our analysis aims at expanding and optimizing the operating domains in CMB-LSS 
cross-correlation studies, similarly to CMB needlet data analysis. It is motivated especially by next 
generation experiments (such as Euclid) which 
will allow us to derive much tighter constraints on cosmological and astrophysical parameters through
cross-correlation measurements between CMB and LSS.}

\begin{document}
\maketitle
\flushbottom

\section{Introduction}
\label{sec:intro}

One of the main puzzles of modern cosmology is the understanding of the mechanism that sources 
the late-time accelerated expansion of the Universe. Whether it is associated to an exotic
form of energy or to some modifications of general relativity, the different scenarios can only be
disentangled by probing the perturbations evolution over cosmic time. 
In this context, galaxy clustering and weak gravitational lensing have become promising probes not only to 
investigate cosmic acceleration but also the dark matter and neutrino sectors. 

While the analysis of the data from the Planck satellite is approaching to an end, yielding a breakthrough in
many respects for what concerns CMB studies \cite{PlanckCollaboration2015c}, such fundamental issues have triggered the 
upcoming experimental efforts and in the next few years
galaxy surveys such as the European Space Agency's (ESA) satellite Euclid\footnote{\url{http://sci.esa.int/euclid/}} \cite{Laureijs2011}, 
the Dark Energy Spectroscopic Instrument (DESI\footnote{\url{http://desi.lbl.gov}}), the Large Synoptic
Survey Telescope (LSST\footnote{\url{http://www.lsst.org}}) and the Wide Field Infrared Survey Telescope
(WFIRST\footnote{\url{http://wfirst.gsfc.nasa.gov}}), along with a plethora of ground-based high-sensitivity 
CMB experiments like the Simons Array\footnote{\url{http://cosmology.ucsd.edu/simonsarray.html}}, the South Pole Telescope (SPT-3G)\footnote{\url{https://pole.uchicago.edu/spt/}}, and the Advanced Atacama 
Cosmology Telescope (AdvACT)\footnote{\url{https://act.princeton.edu}}, will carry out observations devoted to shed light on the physics behind 
the dark components. In these experiments, operating and under design and construction towards the 
efforts of the next decade (including ground-based facilities such as the Simons Observatory\footnote{\url{https://simonsobservatory.org}} and CMB-S4, as well as the 
proposed space satellites COrE\footnote{\url{http://www.core-mission.org}} e LiteBIRD\footnote{\url{http://litebird.jp/eng/}}), the role of CMB-LSS cross correlation is double: on one side, yielding constraints on 
dark energy and matter through the analysis of CMB lensing by forming LSS, and on the other, to de-lens 
the B-modes of polarization in order to improve the constraint, or measure, of the power from primordial 
gravitational waves. 

In particular, LSS data gathered from Euclid in the form of weak lensing and galaxy catalogues will provide 
an excellent tracer for the underlying gravitational potential which is responsible for the CMB lensing effect.
It is then only natural to cross-correlate CMB lensing maps with LSS data to improve the
constraints on dark energy models and cosmological parameters, similarly to what has been done
with CMB temperature and LSS maps in order to extract faint large scale signal like the integrated 
Sachs-Wolfe effect (iSW), see for instance \cite{Pietrobon2006, Vielva2006, McEwen2007, Munshi:2014tua,Ade:2013dsi,Ade:2015dva}.

CMB lensing-galaxy cross-correlation measurements have found different applications in cosmology, such as the reconstruction of the galaxy bias redshift evolution \cite{Bianchini2015,Allison2015a,Bianchini2016}, the investigation of the growth of structures 
\cite{Giannantonio2016}, and the augmentation of the absolute cosmic shear calibration \cite{Baxter2016}. All 
analyses reported to date have reconstructed the 2-point statistics either in harmonic  or real space.

The optimal power spectrum estimator in harmonic space for auto and cross-correlation in presence of mask and anisotropic noise is well known \cite{Tegmark1997} and was used for cross-correlation analysis in \cite{Smith2007} and \cite{Schiavon2013}. Modern iterative algorithms make the exact optimal estimation computationally feasible; despite being potentially suboptimal in cases of very small $f_{\rm sky}$ and highly non-uniform noise, the computational convenience of a fast pseudo-$C_\ell$ (PCL) estimator remains an important property, especially when a cross-correlation analysis must be implemented on a variety of different masks, due to different observational strategies in multiple experiments. This is especially relevant when cross-correlating lensed CMB maps with LSS data. Note that most analysis with very small $f_{\rm sky}$ have been so far performed in the flat-sky approximation, in conjunction with the MASTER algorithm, and they provide a nearly optimal power spectrum estimation. 

In this paper, we shall use instead a procedure based on a wavelet-domain approach; more precisely, we shall discuss how to modify the PCL algorithm to perform a needlet cross-correlation analysis.
Since needlet transform is linear in the data, it cannot perform better than the optimal estimator; however it can improve the performance of linear estimators in the presence of masks, as we shall discuss below. At the same time, a needlet estimator mantains the computational convenience of a  nearly-optimal PCL estimator, provided that the noise properties are fairly uniform.

As discussed in many previous references, needlets are a form of spherical wavelets which were 
introduced in functional analysis and statistics by \cite{Narcowich2006,Baldi2009a} and have then found a 
number of different applications in the cosmological community over the last decade; we recall for 
instance \cite{Marinucci2007} for a general description of the methods, 
\cite{Lan2008,Rudjord2009a,Pietrobon2010a,Donzelli2012,Regan2015,Ade:2015ava} for non-Gaussianity 
estimation, \cite{Delabrouille2010,2014A&A...571A..12P,Adam:2015tpy,Rogers2016,Rogers2016a} for 
foreground component separation, \cite{Geller2008,Leistedt2015a,Ade:2015ava} for polarization data 
analysis, 
\cite{Durastanti2014,Leistedt2015} for extension in 3d framework and \cite{Troja2014,Regan2015} for 
trispectrum analysis.

The advantages of needlets, like those of other wavelets system, have been widely discussed in the 
literature; in short, they are mainly concerned with the possibility to exploit double localization properties, 
in the real and harmonic domain. Despite this localization in the real domain, we show here that the 
performance of a needlet cross-correlation estimator deteriorates badly in the presence of very aggressive 
sky-cuts (i.e., experiments with sky coverage much smaller than 50\%). In this paper, we show how the 
performance of this estimator can be restored by a MASTER-like correction. Thus achieving 
signal-to-noise figure of merits which are in some aspect superior to the corresponding results for power 
spectrum methods; the terms of this comparison are explained in more details below. 

The plan of the paper is as follows. In section \ref{sec:theo} we review quickly some background material 
on both harmonic and needlet cross-correlation analysis; we then proceed in section \ref{sec:master} to 
introduce the MASTER-like algorithm for the needlets cross-correlation estimator. Numerical evidence and 
some comparison on the performance of these procedures are collected in section \ref{sec:num_ev}, 
while final considerations are presented in section \ref{sec:conclusion}.

\section{Building the cross-correlation estimators}
\label{sec:theo}
In this section we introduce the CMB lensing-galaxy cross-correlation estimators in harmonic and 
needlet space.
We start by briefly reviewing the theoretical framework that we exploit to model such signal and show
how the weak lensing of the CMB is correlated to the large scale matter distribution. Then, starting from
these concepts, we illustrate how to build an harmonic estimator of the CMB lensing-galaxy 
cross-correlation signal and finally we derive a needlet estimator from the harmonic one. Even though
the main focus of the paper is the measurement of the CMB lensing-galaxy cross-correlation,
we recall that the estimators presented here can be applied to any scalar field on the sphere.

\subsection{The weak lensing of the CMB from LSS}
\label{sec:xcorr-need-theo}

Gravitational lensing performs a remapping of the primordial CMB temperature and polarization 
anisotropies by a deflection
field $\mathbf{d}(\nver)$, so that photons coming from direction $\nver$ carry information about the patch 
of the sky in the perturbed direction $\nver + \mathbf{d}(\nver)$ (see \cite{Lewis2006}). 
The deflection field can be written as the gradient of a scalar potential, namely
the CMB lensing potential $\phi(\nver)$, which encodes information about the Weyl potential
\footnote{Here we define the Weyl potential as $(\Psi+\Phi)/2$, half the sum of the two Bardeen potentials 
$\Psi$ and $\Phi$.}
integrated out to the last-scattering surface. Here we work in terms of the (spherical) Laplacian of the 
lensing potential,
the CMB convergence field\footnote{This relation translates in harmonic space into 
$\kappa_{\ell m} = \frac{\ell(\ell+1)}{2}\phi_{\ell m}$.}  $\kappa(\nver) = -\Delta_{S^2}\phi(\nver)/2$ which 
describes the local (de)magnification of CMB fluctuations, while the Laplacian reads as 
$\Delta_{S^2}=\frac{1}{\sin\theta}\frac
{\partial}{\partial\theta}
\left(\sin\theta\frac{\partial}{\partial\theta}\right)+\frac{1}{\sin^2\theta}\frac{\partial^2}{\partial\varphi^2}$. As concerns the tracer galaxies we define the projected galaxy density fluctuations 
as $g(\nver)= n(\nver)/\bar{n}-1$, where $n(\nver)$ is the number of objects in a given direction, and $\bar{n}$ is the mean
number of sources. In standard cosmologies both the CMB convergence and the galaxy overdensity can be written as a weighted 
integral of the matter overdensity $\delta(\nver)$ along the line-of-sight (LOS):

\begin{equation}
X(\nver) = \int_0^{z_*} dz\, W^X(z)\delta(\chi(z)\nver,z),
\end{equation}
where $X=\{\kappa,g\}$ and $W^X(z)$ is the kernel related to a given field.\\
The kernel $W^{\kappa}$ quantifies the matter distribution lensing efficiency and it reads
\begin{equation}
W^{\kappa}(z) = \frac{3\Omega_m}{2c}\frac{H_0^2}{H(z)}(1+z)\chi(z)\frac{\chi_*-\chi(z)}{\chi_*}.
\end{equation}
Here $H(z)$ is the Hubble factor at redshift $z$, $\chi(z)$ is the comoving distance to redshift $z$, $\chi_*$ is the comoving distance to the last scattering surface at $z_*\simeq
1090$, $c$ is the speed of light, $\Omega_m$ and $H_0$ are the present-day values of matter
density and Hubble parameter, respectively.\\
Under the hypothesis that luminous matter traces the peaks of the underlying dark matter field, we write the observed projected galaxy overdensity as the sum of an intrinsic clustering term and a lensing magnification bias one, so that the galaxy kernel reads
\begin{equation}
W^{g}(z) = b(z)\frac{dN}{dz} + \mu(z).
\label{eqn:wg}
\end{equation}
The former term describes the physical clustering of the sources and is given by the product of the bias factor $b$ with the \emph{unit-normalized} redshift distribution of galaxies, $dN/dz$. The latter is related to the lensing magnification bias and it writes:
\begin{equation}
\label{eqn:wmu}
\mu(z) = \frac{3\Omega_{\rm m}}{2c}\frac{H_0^2}{H(z)}(1+z)\chi(z) \int_z^{z_*}dz'\,\left(1-\frac{\chi(z)}{\chi(z')}\right)(\alpha(z')-1)\frac{dN}{dz'}.
\end{equation}
Magnification bias is independent of the tracer bias parameter and, in the weak lensing limit, depends on the slope of the galaxy number counts $\alpha$ ($N(>S)\propto S^{-\alpha}$) at the flux density limit of the survey. \\
At smaller angular scales ($\ell \gtrsim 20$), the Limber approximation \cite{Limber1953} allows us to relate the theoretical two-point statistics of the CMB convergence-galaxy and galaxy-galaxy correlations to the matter power spectrum $P_{\delta\delta}(k,z)$ through:
\begin{equation}\label{eq:cross}
\begin{split}
C_{\ell}^{\kappa g} &=   \int_0^{z_*} \frac{dz}{c} \frac{H(z)}{\chi^2(z)} W^{\kappa}(z)W^{g}(z)P_{\delta\delta}\biggl(\frac{\ell}{\chi(z)},z\biggr); \\
C_{\ell}^{gg} &=   \int_0^{z_*} \frac{dz}{c} \frac{H(z)}{\chi^2(z)} [W^{g}(z)]^2P_{\delta\delta}\biggl(\frac{\ell}{\chi(z)},z\biggr).
\end{split}
\end{equation}
We calculate the matter power spectrum using the \texttt{CAMB}\footnote{\url{http://cosmologist.info/camb/}} code \cite{Lewis2000},
including the effect of non-linear matter clustering via the common \texttt{Halofit} prescription \citep{Takahashi2012}.

\subsection{Harmonic cross-correlation estimator}
\label{subsec:stdcorr}
Most of the cosmological observations, from CMB to galaxy surveys, provide us with data in the form 
of two-dimensional sky-maps.\footnote{This especially applies when distance information about the sources is unavailable, nevertheless the quantity of interest can always be projected on the sphere.} 
The information content hidden in such maps is usually probed by means of harmonic analysis on the sphere. A popular observable that characterizes the statistical properties of a given cosmic field is the
angular power spectrum $C_{\ell}$ and its reconstruction enables a direct comparison between
models and data.

It is common practice to decompose the observed field $X(\nver)$ into spherical harmonics, a frequency-space orthonormal basis for representing functions defined over the sphere, as
\begin{equation}
X(\nver) = \sum_{\ell m} x_{\ell m}Y_{\ell m}(\nver),
\label{eqn:spharm}
\end{equation}
where the spherical harmonic coefficients are given by
\begin{equation}
x_{\ell m} = \int_{\mathbb{S}^2}X(\nver)Y^*_{\ell m}(\nver)d\Omega.
\label{eqn:sphcoeff}
\end{equation}
For an isotropic finite variance field we have that the mean of the spherical harmonic coefficients is $\langle x_{\ell m}\rangle = 0$,
while their covariance is given by $\langle x_{\ell m} x^*_{\ell' m'}\rangle = C^{XX}_{\ell}\delta_{\ell\ell'}\delta_{mm'}$.
An unbiased estimator of the (cross-)angular power spectrum is given by (hereafter the hat $\hat{X}$ 
denotes  estimated quantities):
\begin{equation}
\hat{C}_{\ell}^{XY} = \frac{1}{2\ell+1}\sum_{m=-\ell}^{\ell} x_{\ell m}y_{\ell m}^*.
\label{eqn:sphest}
\end{equation}
In particular, $\hat{C}_{\ell}$ can be shown to possess the minimal variance among the unbiased 
estimators (in the sense that its variance reaches the Cram\`er-Rao lower bound) for spatially uniform noise and in the absence of mask (see \cite{Tegmark1997}).

Spherical harmonics are particularly appealing because they are statistically orthogonal for full-sky 
Gaussian-distributed sky-maps, i.e. the covariance is diagonal $\text{Cov}_{\ell \ell'}\propto \delta_{\ell
\ell'}$, and the power spectrum fully characterizes the behaviour of the field. However, real-world observations have to deal with
a number of limitations and issues, such as the finite instrumental spatial resolution, the anisotropic 
noise, and asymmetric beam response. Moreover the incomplete sky coverage, motivated for example 
by foreground contamination or the instruments scanning strategy, induces a mode-coupling and a power leakage between 
different multipoles, as well as an overall downward shift of power \citep{Hivon2001,Efstathiou2004}. This makes the exact 
evaluation of the following spherical harmonic transform cumbersome:
\begin{equation}
\begin{split}
\tilde{x}_{\ell m} &= \int_{\mathbb{S}^2}X(\nver)W(\nver)Y^*_{\ell m}(\nver)d\Omega \\
&= \sum_{\ell' m'} K_{\ell m \ell' m' }[W] x_{\ell' m'},
\end{split}
\label{eqn:sphcoeffmask}
\end{equation}
where the kernel $K$, dependent on the weighting scheme $W(\nver)$ (i.e. the mask), describes the induced mode-coupling. 
A common approach to obtain unbiased but slightly sub-optimal bandpower estimates is to use approximate PCL 
methods as the well known MASTER (see i.e., \citep{Hivon2001,Efstathiou2004}). It is possible to show that the pseudo-spectrum 
$\tilde{C}^{XY}_{\ell} = (2\ell+1)^{-1}\sum_m \tilde{x}_{\ell m}\tilde{y}^*_{\ell m}$ is related to the underlying power spectrum $C_{\ell}$ as 
\begin{equation}
\label{eqn:pcl2cl}
\langle \tilde{C}^{XY}_{\ell}\rangle = \sum_{\ell'} M_{\ell\ell'} C^{XY}_{\ell'},
\end{equation}
where $M_{\ell\ell'}$ is the coupling matrix as defined in \cite{Hivon2001}. The basic idea is to invert eq.~(\ref{eqn:pcl2cl}) in order to 
recover the underlying power spectrum, however for small sky fraction  $f_{\rm sky}=\frac{1}{4\pi}\int_{\mathbb{S}^2}W^2(\nver)d\Omega$, 
one needs to bin the pseudo-power spectrum and the coupling matrix, so that the estimator of the true
 cross-bandpowers $\hat{C}^{XY}_{L}$ writes
\begin{equation}
\label{eqn:master_xy}
\hat{C}^{XY}_{L} = \sum_{L' \ell}K^{-1}_{LL'}P_{L'\ell}\tilde{C}^{XY}_{\ell},
\end{equation}
where $L$ is the bandpower index and the binned coupling matrix can be written as
\begin{equation}
K_{LL'} = \sum_{\ell\ell'} P_{L\ell}M_{\ell\ell'}B^2_{\ell'}Q_{\ell' L'}.
\end{equation}
Here $P_{L\ell}$ is the binning operator, $Q_{\ell L}$ is its reciprocal, and $B^2_{\ell'}$ is the pixel window function that corrects for the finite pixel size. If the true power spectrum varies slowly with respect to the coupling matrix and/or $f_{\rm sky}$ is large, eq.~(\ref{eqn:pcl2cl}) becomes  
\begin{equation}
\label{eqn:fsky_app}
\langle \tilde{C}^{XY}_{\ell}\rangle \approx  C^{XY}_{\ell}  \sum_{\ell'} M_{\ell\ell'} = f_{\rm sky} C^{XY}_{\ell},
\end{equation}
which is the so-called $f_{\rm sky}$ approximation \citep{Komatsu2002a}.

\subsection{Needlet cross-correlation estimator}
\label{subsec:needcorr}
As mentioned in the introduction, some drawbacks of standard Fourier analysis on the sphere can be mitigated by the exploitation of needlet/wavelet techniques. Related advantages have already been widely discussed in the literature, see again \cite{Marinucci2007,Lan2008,Donzelli2012,Troja2014,Marinucci2011,Durastanti2014}.

Here we simply recall that the spherical needlet system $\psi_{\{jk\}}$ can be obtained by a quadratic combination of spherical harmonics as
\begin{equation}
\label{eqn:needfunc}
\psi_{jk}(\nver) = \sqrt{\lambda_{jk}} \sum_{\ell=[B^{j-1}]}^{[B^{j+1}]} b\biggl(\frac{\ell}{B^{j}}\biggr)\sum_{m=-\ell}^{\ell}Y^*_{\ell m}(\nver)Y_{\ell m}(\xi_{jk}),
\end{equation}
where $[\cdot]$ denotes the integer part, $b(\cdot)$ is the filter function in the harmonic domain defined for $x\in [1/B,B]$, and $\{\xi_{jk}\}$ are the cubature points on the sphere corresponding to the frequency $j$ and the location $k$. Since our implementation relies on the \texttt{HEALPix}\footnote{\url{http://healpix.jpl.nasa.gov}} \cite{Gorski2005}
 pixelation scheme we can identify the cubature points with the pixel centers, so that the cubature weights $\lambda_{jk}$ can be approximated by $4\pi/N_{\rm pix}$, where $N_{\rm pix}$ is the number of pixels for the chosen \texttt{HEALPix} $N_{\rm side}$ resolution and $k$ represents the pixel number \cite{Pietrobon2006}. 
 
\begin{figure}[tbp]
\centering 
\includegraphics[width=0.6\textwidth]{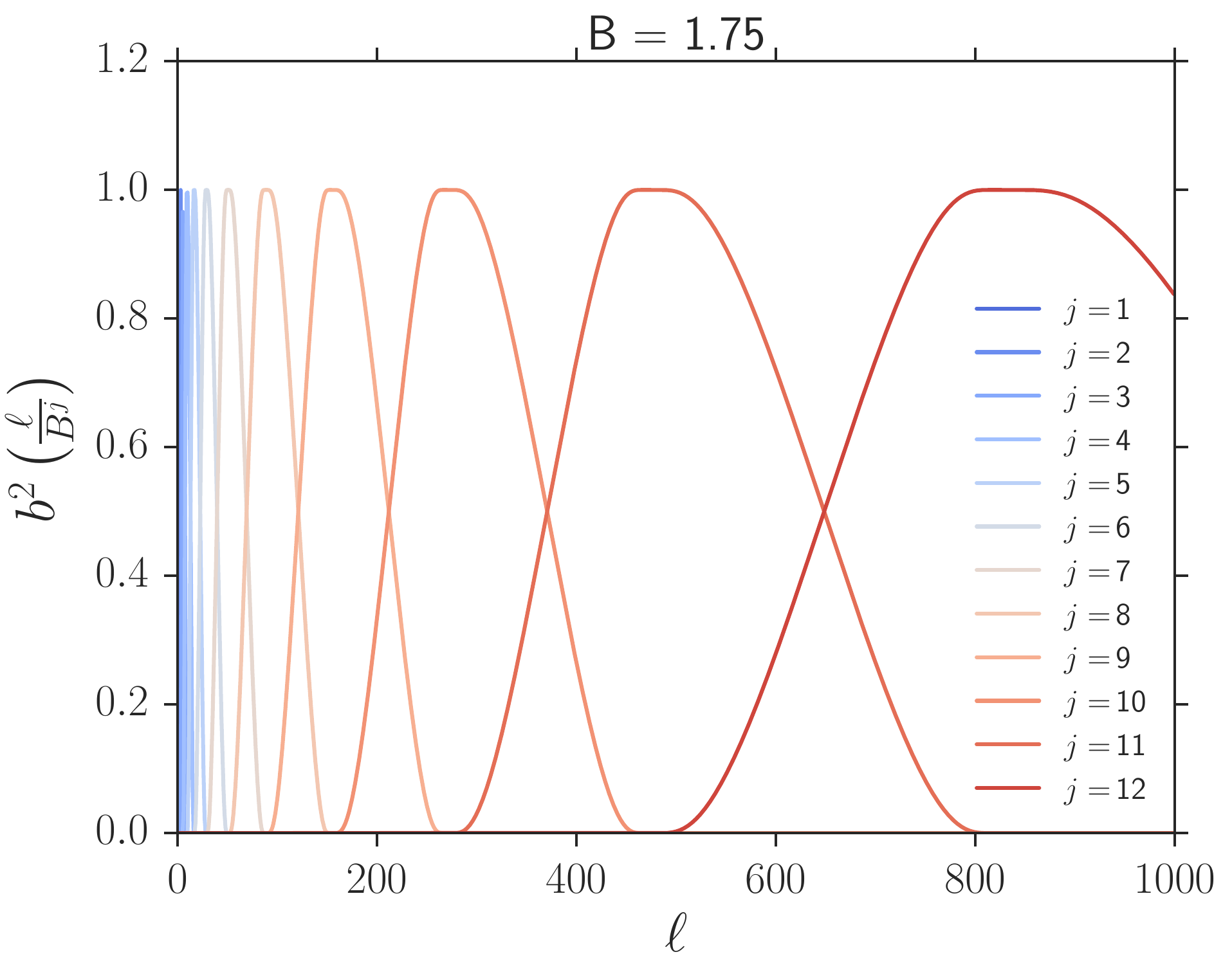}
\caption{\label{fig:b_need} Profile of the filter function in the $\ell$-space for different needlet frequencies $j$. The needlet width parameter is set to $B=1.75$.}
\end{figure}

Needlets can be thought of as a convolution of the projection operator $\sum_m Y^*_{\ell m}(\nver)Y_{\ell m}(\xi_{jk})$ with a filter function $b(\cdot)$ whose width is controlled by the only free parameter $B$: recipes for the construction of the function $b(\cdot)$ can be found in \cite{Marinucci2007,McEwen2013,Marinucci2011}. A smaller value of $B$ corresponds to a narrower localization in $\ell$-space, while a larger value translates into a more precise localization in real space. Once $B$ is fixed, each needlet can be shown to pick up signal only from a specific range of multipoles determined by the index $j$: the profile of the filter function $b(\cdot)$ is shown in figure~\ref{fig:b_need} for different frequencies.
Needlet coefficients are then evaluated by projecting the centered field $X(\nver)$ on the corresponding needlet $\psi_{jk}(\nver)$ as
\begin{equation}
\label{eqn:needcoef}
\begin{split}
\beta_{jk} &= \int_{\mathbb{S}^2} X(\nver)\psi_{jk}(\nver)d\Omega\\
&= \sqrt{\lambda_{jk}} \sum_{\ell=[B^{j-1}]}^{[B^{j+1}]} b\biggl(\frac{\ell}{B^{j}}\biggr)\sum_{m=-\ell}^{\ell}x_{\ell m}Y_{\ell m}(\xi_{jk}).
\end{split}
\end{equation}
Needlet coefficients corresponding to a given frequency $j$ can themselves be represented as an 
\texttt{HEALPix} map. It is worth to stress that although needlets do not
make up an orthonormal basis for square integrable functions on the sphere, they represent a tight frame (redundant basis) so that they allow for a
simple reconstruction formula. After computing the needlet coefficients $\beta_{jk}$ from the maps, we can build a spectral estimator as
\begin{equation}
\label{eqn:needest}
\hat{\beta}^{XY}_{j} = \frac{1}{N_{\rm pix}} \sum_k \beta^X_{jk}\beta^Y_{jk},
\end{equation}
and it is immediate to check that it provides an unbiased estimate of (a binned form of) the angular power spectrum, i.e.
\begin{equation}
\label{eqn:needestmean}
\langle \hat{\beta}^{XY}_{j} \rangle \equiv \beta^{XY}_{j} = \sum_{\ell} \frac{2\ell+1}{4\pi}  b^2 \biggl( \frac{\ell}{B^j} \biggr) C^{XY}_{\ell}.
\end{equation}
These theoretical predictions can directly be compared to the extracted spectra, allowing for the parameter extraction process.
Moreover, as noted in \cite{Pietrobon2006}, the analytic relation between $\beta_j$ and $C_{\ell}$ makes straightforward dealing 
with beam profiles, pixel window function, and experimental transfer functions. 
Note that in this paper we divide the spectral estimator \ref{eqn:needest} and its expected value \ref{eqn:needestmean} for a normalizing
factor $\mathcal{N}$ given by 
\begin{equation}
\label{eqn:norm}
\mathcal{N} = \sum_{\ell} \frac{2\ell+1}{4\pi}  b^2\biggl( \frac{\ell}{B^j} \biggr),
\end{equation}
so that in the plots we show $\hat{\beta}^{XY}_{j} \to \hat{\beta}^{XY}_{j}/\mathcal{N}$.\\
The theoretical variance of the cross-correlation power spectrum in needlet space reads
\begin{equation}
\label{eqn:needestvar}
(\Delta\beta^{XY}_j)^2 \equiv \text{Var}[\hat{\beta}^{XY}_j]= \sum_{\ell} \frac{2\ell+1}{16\pi^2}  b^4 \biggl( \frac{\ell}{B^j} \biggr)
 \bigl[ (C_{\ell}^{XY})^2 + C_{\ell}^{XX}C_{\ell}^{YY} \bigr],
\end{equation}
where the angular auto-spectra can be comprehensive of a noise term, i.e. 
$C_{\ell} \to C_{\ell} + N_{\ell}$, if present. Moreover, the needlets system 
is compactly supported in the harmonic domain and as such, for full-sky maps, the random needlets coefficients are uncorrelated by 
construction for $|j-j'| \ge 2$ \citep{Baldi2009a}.

\section{MASTER algorithm for needlets}
\label{sec:master}

As mentioned in Sec~\ref{subsec:needcorr}, one of the main driver behind the development of the needlet spectral estimator is 
the need to overcome the issues related to Fourier analysis on the sphere in the presence of missing observation.
The excellent needlets localization properties in real space represent a key feature for analyzing cosmological data on the 
partially observed sky, in particular it has been shown that even in the presence of masked regions the random needlet coefficients 
$\beta_{jk}$ are asymptotically independent (over $k$) as $j \to \infty$ (contrary to the case of random coefficients $x_{\ell m})$ 
\citep{Marinucci2007, Baldi2009a}. However, as we shall see from simulations in the next section, the estimator defined in eq.~(\ref{eqn:needest})
becomes biased for aggressive masking: here we formally study the effect of sky-cuts on the needlet power spectrum estimation.\\ 
From eq.~(\ref{eqn:sphcoeffmask}), we find that needlet coefficients computed on a masked sky are given by
\begin{equation}
\label{eqn:needcoefmask}
\tilde{\beta}_{jk} = \sqrt{\lambda_{jk}} \sum_{\ell} b\biggl(\frac{\ell}{B^{j}}\biggr)\sum_{m}\tilde{x}_{\ell m}Y_{\ell m}(\xi_{jk}).
\end{equation}
Then, if we consider the statistic 
\begin{equation}
\label{eqn:gammaest}
\hat{\Gamma}^{XY}_j = \frac{1}{N_{\rm pix}} \sum_k \tilde{\beta}^X_{jk} \tilde{\beta}^Y_{jk},
\end{equation}
it is straightforward to see that its expectation value reads as follows
\begin{equation}
\label{eqn:gammaestmean}
\begin{split}
\langle \hat{\Gamma}^{XY}_j \rangle \equiv {\Gamma}^{XY}_j &= \sum_{\ell m} b^2\biggl(\frac{\ell}{B^{j}}\biggr) \langle \tilde{x}_{\ell m}\tilde{y}_{\ell m} \rangle \\
&= \sum_{\ell m} \sum_{\ell' m'} b^2\biggl(\frac{\ell}{B^{j}}\biggr) K^2_{\ell m \ell' m'}[W] C_{\ell'}\\
&= \sum_{\ell\ell'} \frac{2\ell+1}{{4\pi}} b^2\biggl(\frac{\ell}{B^{j}}\biggr) M_{\ell\ell'}C_{\ell'},
\end{split}
\end{equation}
which tells us that $\hat{\Gamma}^{XY}_j $ is an unbiased estimator for a smoothed version of the pseudo angular power spectrum $\tilde{C}_{\ell}$,
similar to the case of $\hat{\beta}^{XY}_j$: in some sense, we can view $\hat{\Gamma}^{XY}_j $ as an estimator of the pseudo-needlet power spectrum. Using eq.~(\ref{eqn:fsky_app}), which is valid for slowly varying power spectra and/or large sky fractions,
it is possible to relate the two estimators as 
\begin{equation}
\label{eqn:gammabeta}
\begin{split}
\langle \hat{\Gamma}^{XY}_j \rangle &\approx  f_{\rm sky}\sum_{\ell} \frac{2\ell+1}{{4\pi}} b^2\biggl(\frac{\ell}{B^{j}}\biggr) C_{\ell}\\
&= f_{\rm sky} \langle \hat{\beta}^{XY}_j \rangle.
\end{split}
\end{equation}
Before we conclude this section let us introduce a couple of remarks. We recall first that in the case of a survey with a large sky-cut, inverting the full coupling matrix become unfeasible because of singularities; hence the power spectrum can be estimated only over some subset of multipoles i.e. the power spectrum is recovered only up to some frequency windows.
As discussed earlier in Sec~\ref{subsec:stdcorr}, the choice of this frequency windows is to a good degree arbitrary; on the other hand, the needlet framework naturally provides a binning 
scheme which is  controlled by a single width parameter $B$ (as well as by the profile of the filter function $b(\cdot)$).

As a second difference, we note that while the  PCL approach  usually
makes use of the \emph{backward modelling}, where measurements are deconvolved for numerical and observational 
effects to match the theoretical predictions, needlets analysis is oriented towards the 
\emph{forward modelling}, which turns theoretical (needlet) power spectra into pseudo-spectra that can be directly compared to the raw 
measurements\footnote{Note that pseudo-spectra, either in harmonic or needlet space, depend on the observational setup represented
for example by the masking, the smoothing, and the apodization, while this is not the case for theoretical predictions.} (see \cite{Harnois-Deraps2016} 
for a  discussion on forward and backward modeling). 
In particular, in the needlet case it is not feasible to write a closed formula such as eq.~(\ref{eqn:master_xy}) to express the original needlet power spectrum as a function of the pseudo one, i.e. $\beta_j = \beta_j(\Gamma_j)$; however, this is not an obstacle for data analysis because the forward estimator can be used just as well to do model checks as parameter estimation. In particular, there does not seem to be any intrinsic advantage by using either backward or forward modelling in terms of signal-to-noise-ratio.

\begin{figure}[tbp]
\centering 
\includegraphics[width=0.8\textwidth]{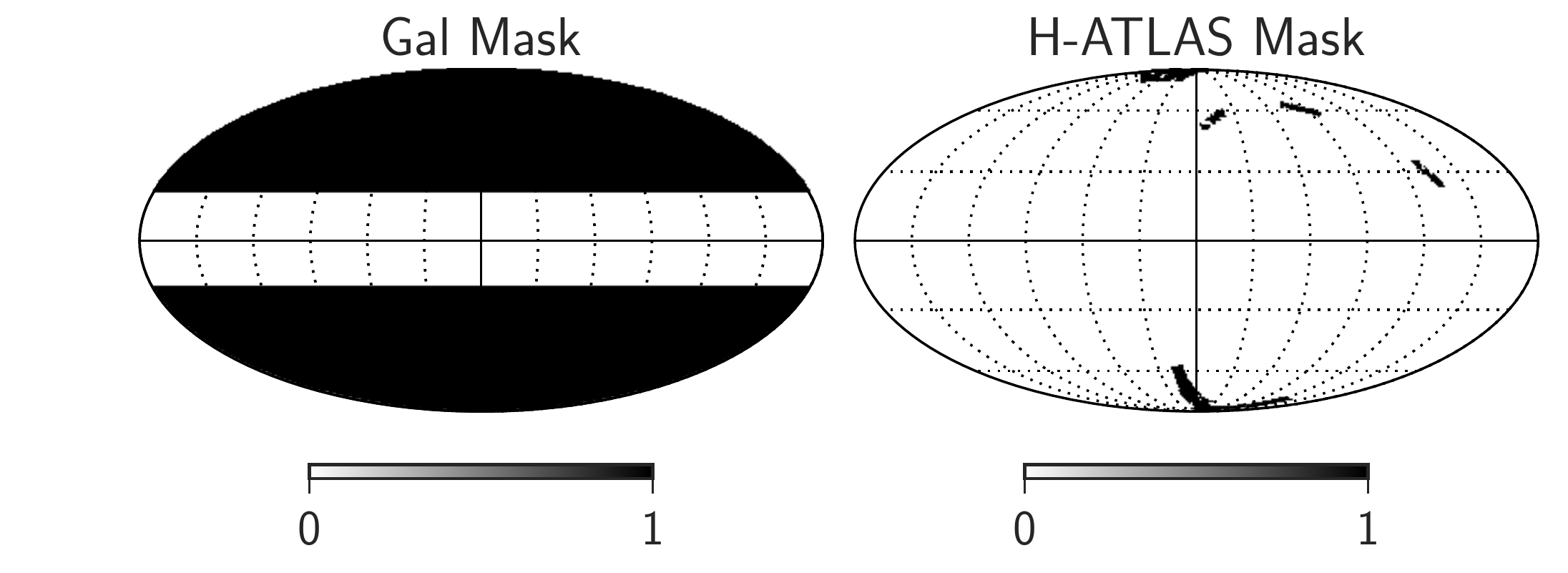}
\caption{\label{fig:masks} Masks used for the analysis. The mask with a symmetric galactic cut at $\pm 20\deg$ ($f_{\rm sky}= 0.65$) is shown
in the left part, while the H-ATLAS mask ($f_{\rm sky} = 0.013$) is shown on the right one. In both cases the black color denotes
observed regions of the sky.}
\end{figure}

\section{Numerical evidence}
\label{sec:num_ev}
In this section we describe the simulations setup exploited and the tests performed in order to compare the harmonic and needlet cross-correlation estimators.

\subsection{Simulations}
\label{sec:sims}
We simulate a set of $N_{\rm sim}=500$ correlated CMB convergence and galaxy density maps at an 
HEALPix resolution of $N_{\rm side} = 512$ (corresponding to an angular resolution of $\sim 7'.2$). 
For the galaxies we consider an high-$z$ Herschel-like 
population with a redshift distribution as described in \cite{Bianchini2016} and fix $b=3$ for the present 
galaxy sample; the precise details of spectra are not fundamental since we are interested in testing the 
estimators. This simulations set is used in a Monte Carlo (MC) 
approach (i) to validate the extraction pipelines; (ii) to compute the uncertainty associated with each bin; 
and (iii) to quantify the degree of correlation among different needlet frequencies. A thorough description 
of the main steps to obtain correlated CMB lensing and galaxy maps comprising of signal and noise can 
be found in \cite{Bianchini2015}, here we simply use noise-free maps for validation purposes. 
Pairs of correlated signal-only Gaussian CMB convergence $\kappa^S_{\ell m}$ and galaxy density 
$g^S_{\ell m}$ maps are generated from the three fiducial spectra $C^{\kappa g}_{\ell}$, $C^{\kappa
\kappa}_{\ell}$ and $C^{gg}_{\ell}$ \cite{Giannantonio2008,Bianchini2015}. This is easily implemented 
using the synfast routine of HEALPix.
In order to show the effect of masking on the reconstructed statistics, the simulated maps are
masked with two different masks: we consider either a galactic mask that covers the 35\% of sky 
($f_{\rm sky} = 0.65$), similar to the one implemented in Planck CMB data, and a much more 
aggressive Herschel-Astrophysical Terahertz Large Area Survey (H-ATLAS) \cite{Eales2010a} mask with 
sky coverage equal only to 1.3\% that comprehends the North Galactic Pole, the South Galactic Pole, and 
the GAMA fields. The adopted masks are shown in figure~\ref{fig:masks}.

\subsection{Results}
\label{sec:results}

In this subsection we present and discuss the tests performed in order to compare the harmonic and needlet cross-correlation estimators. For the former estimator we reconstruct the angular power spectrum in 39 linearly-spaced bandpowers between $\ell \in [2,782]$ with a bin size of $\Delta\ell=20$, while for the latter we fix $B=1.75$ and consider a maximum needlet frequency of $j_{\rm max}=12$.

We start by investigating the uncorrelation properties of the needlets coefficients, as a function of the width of the mask. In particular, the covariance matrix of needlet coefficients is computed by means of 500 MC simulations as 
\begin{equation}
\label{eq:fullcov}
\text{Cov}_{jj'} \equiv \text{Cov}[\hat{\beta}_{j},\hat{\beta}_{j'}] = \langle(\hat{\beta}_j-\langle \hat{\beta}_j\rangle_{\rm MC})(\hat{\beta}_{j
'}-\langle\hat{\beta}_{j'}\rangle_{\rm MC}) \rangle_{\rm MC}.
\end{equation}

\begin{figure}[H]
\centering 
\includegraphics[width=0.8\textwidth]{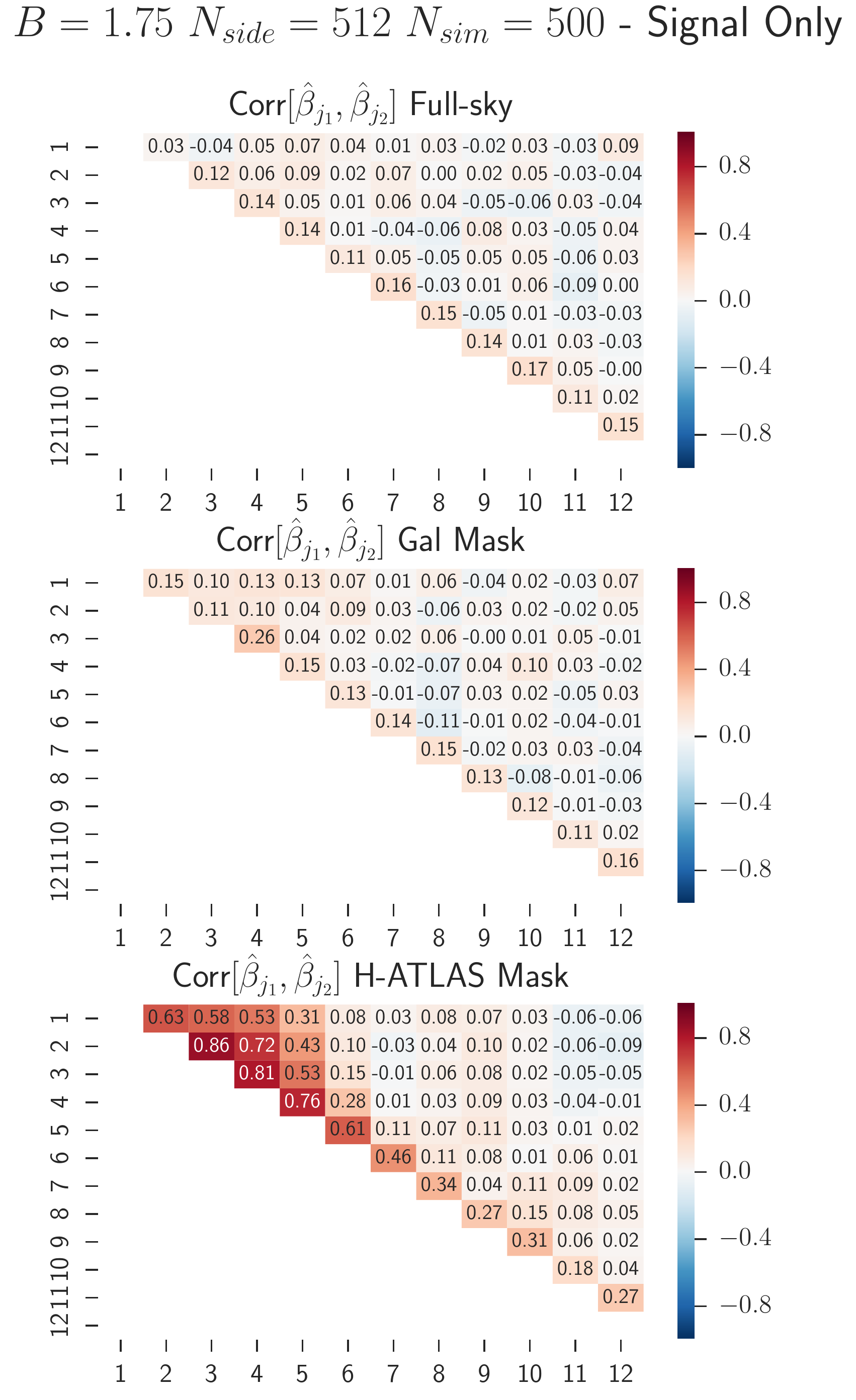}
\caption{\label{fig:needsims_corr} Cross-correlation coefficient matrices, defined as $\text{Corr}_{ij} \equiv \text{Cov}_{ij}/\sqrt{\text{Cov}_{ii}\text{Cov}_{jj}}$, about the needlet space 
estimator. From top to bottom we show results for the full-sky, galactic mask, and H-ATLAS mask
cases respectively. }
\end{figure}

\begin{figure}[t]
\centering 
\includegraphics[width=0.49\textwidth]{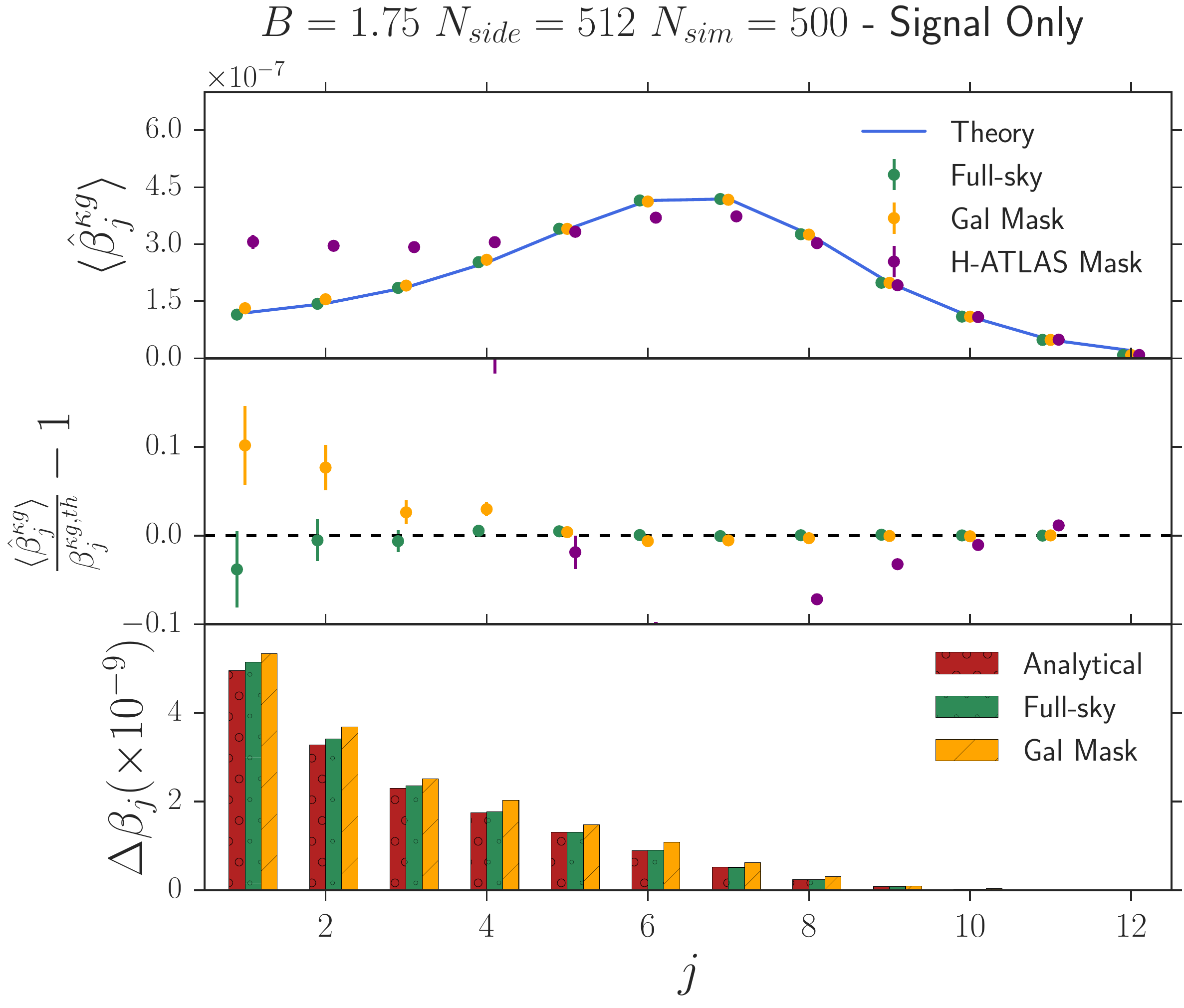}
\includegraphics[width=0.49\textwidth]{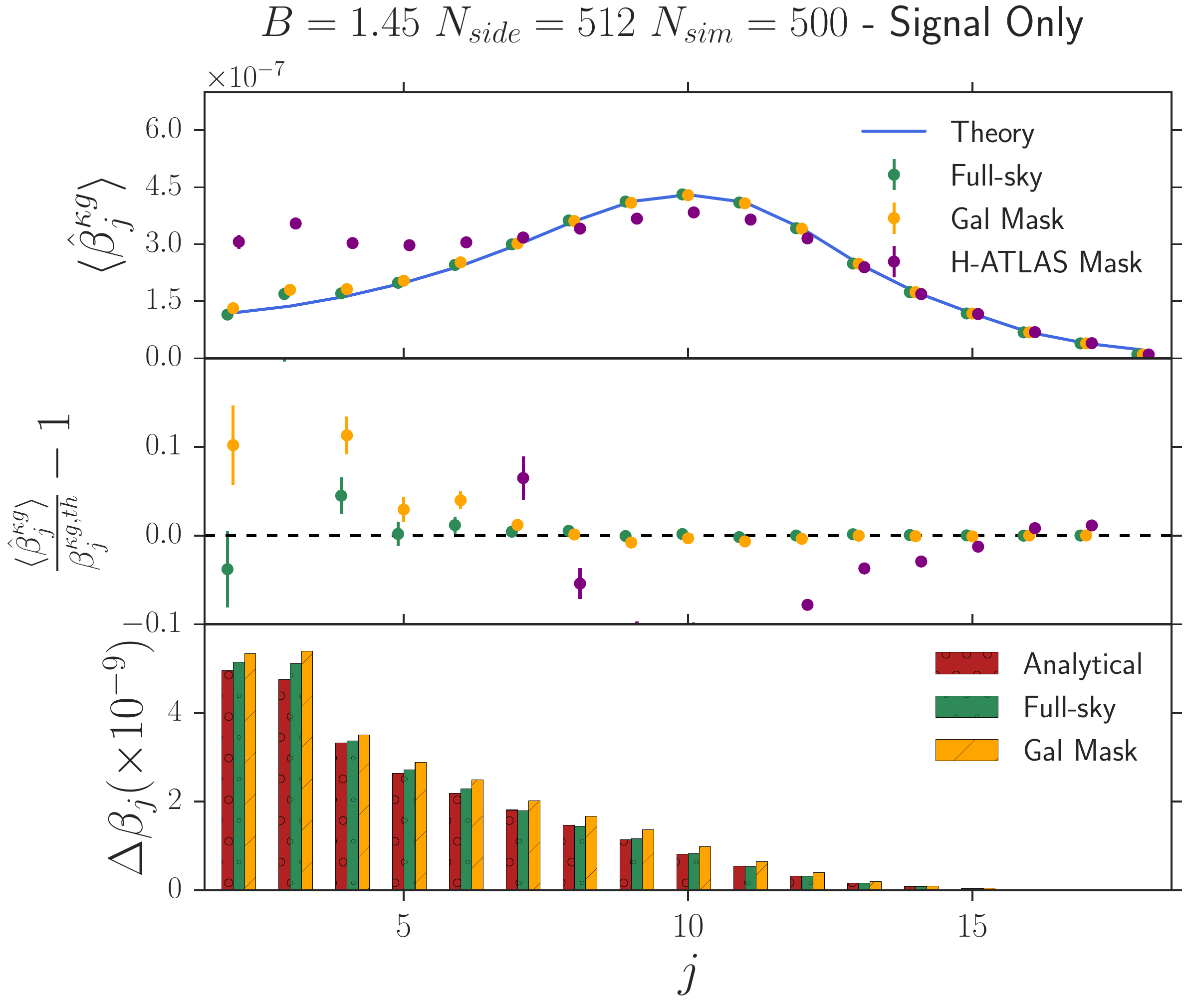}
\caption{\label{fig:needsims} \emph{Upper panel}: Recovered mean needlet cross-power spectrum  
between correlated CMB convergence and galaxy density maps for different masks and width parameter
($B=1.75$ and $1.45$ on the left and right parts respectively). Green, yellow and purple 
bandpowers represent full-sky, galactic mask (with $f_{\rm sky} = 0.65$) and H-ATLAS mask (with $f_{\rm 
sky} = 0.013$) cases respectively. Solid blue line is the generative theoretical input cross-power spectrum. 
Error bars shown are the diagonal components of the covariance matrices (defined in eq. 4.1), properly 
scaled by $\sqrt{N_{\rm sim}}$. Note that reconstructed mean needlet power spectra $\langle \hat{\beta}^{\kappa g}_j \rangle$ are corrected for the observed sky fraction using eq.~(\ref{eqn:gammabeta}). \emph{Central panel}: Fractional difference between mean recovered and 
theoretical needlet cross-spectra for the cases shown in the upper panel. \emph{Lower panel}: Error bars 
comparison for the cases shown in the upper panel. Note that the 
lack of power observed for $j = 12$ (or for $j=18$ if $B=1.45$) is due to the fact that simulated maps have been generated using 
spectral information up to $\ell_{\rm max}=2N_{\rm side}= 1024$, while the needlet frequency $j=12$
picks up signal in the multipole range of $458 \lesssim \ell \lesssim 1396$ ($551 \lesssim \ell \lesssim 1159)$, where the power is partially 
missing.}
\end{figure}

\begin{figure}[H]
    \centering
    \begin{minipage}[t]{0.5\textwidth}
        \centering
        \includegraphics[width=\textwidth]{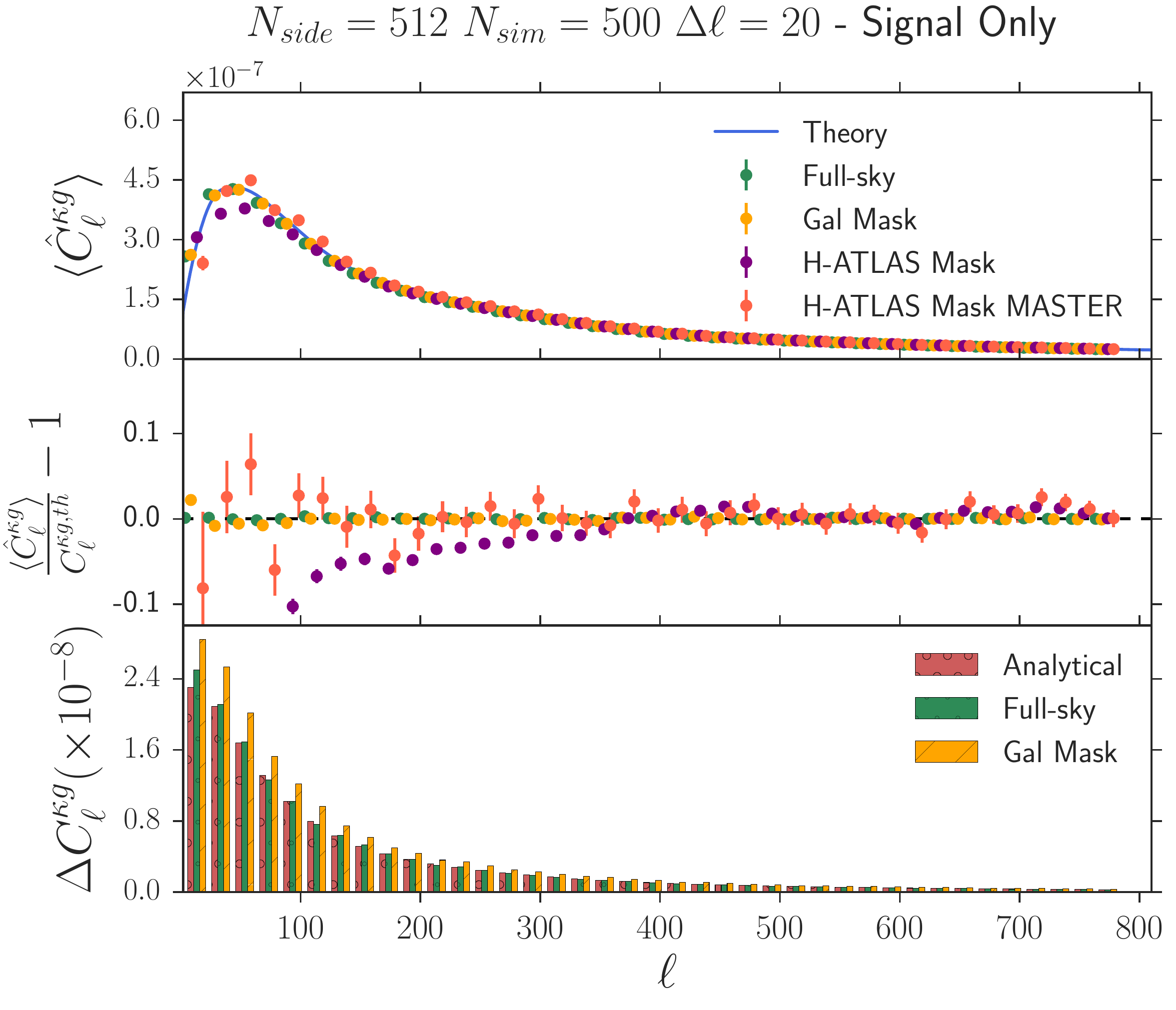}
	\caption{\label{fig:clsims} Same as \ref{fig:needsims} but in harmonic space. }
    \end{minipage}%
    \begin{minipage}[t]{0.5\textwidth}
        \centering
        \includegraphics[width=\textwidth]{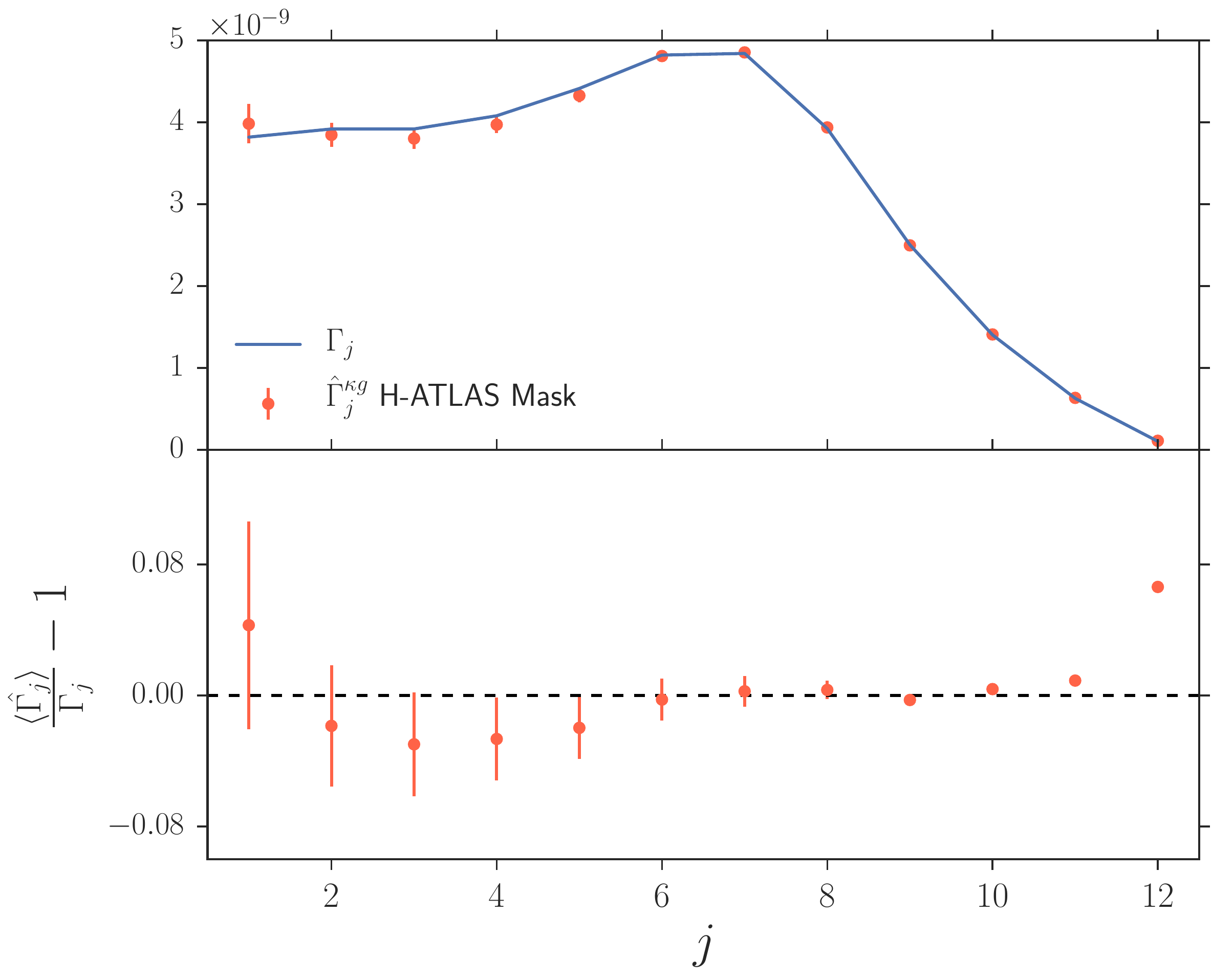}
        \caption{Mean needlet pseudo power spectrum $\langle \hat{\Gamma}^{\kappa g}_j\rangle$ 
        (orange circles) superimposed to the generative theoretical (pseudo) spectrum (blue line).}
        \label{fig:gammaj}
    \end{minipage}
\end{figure}

The corresponding results are reported in figure~\ref{fig:needsims_corr} for the full sky galactic and H-ATLAS case respective. Numerical evidence is very much consistent with the theoretical expectation: in particular in the full sky and galactic mask case the correlation decrease very rapidly outside the main diagonal (where it is trivially unit, which is not reported in the table) in the case of full-sky maps. The decay is still very satisfactory when sky coverage is high, although not complete as for the galactic mask; on the other hand a very aggressive cut with sky coverage of 1.3\% deteriorate enormously the uncorrelation properties (even though Corr$_{jj'}$ is $\mathcal{O}(0.1)$ and smaller at high frequencies for $|j-j'|\ge2$), see the bottom panel in figure~\ref{fig:needsims_corr}. The estimated covariances are then used to derive error bars in the cross-correlation estimators reported in figure~\ref{fig:needsims}. Again, the needlet estimator is shown to perform very well in the full-sky and galactic mask cases whereas Herschel-like framework clearly requires corrections. Error bars decay rapidly for increasing frequencies as expected.
For comparison, in figure~\ref{fig:clsims} MASTER-like estimators are reported for the cross-power spectrum, while the equivalent MASTER needlet reconstruction discussed in section~\ref{sec:master}  is shown in figure~\ref{fig:gammaj}, where we can see that the bias is strongly suppressed. 

\begin{figure}[tbp]
\centering 
  \includegraphics[width=0.49\textwidth]{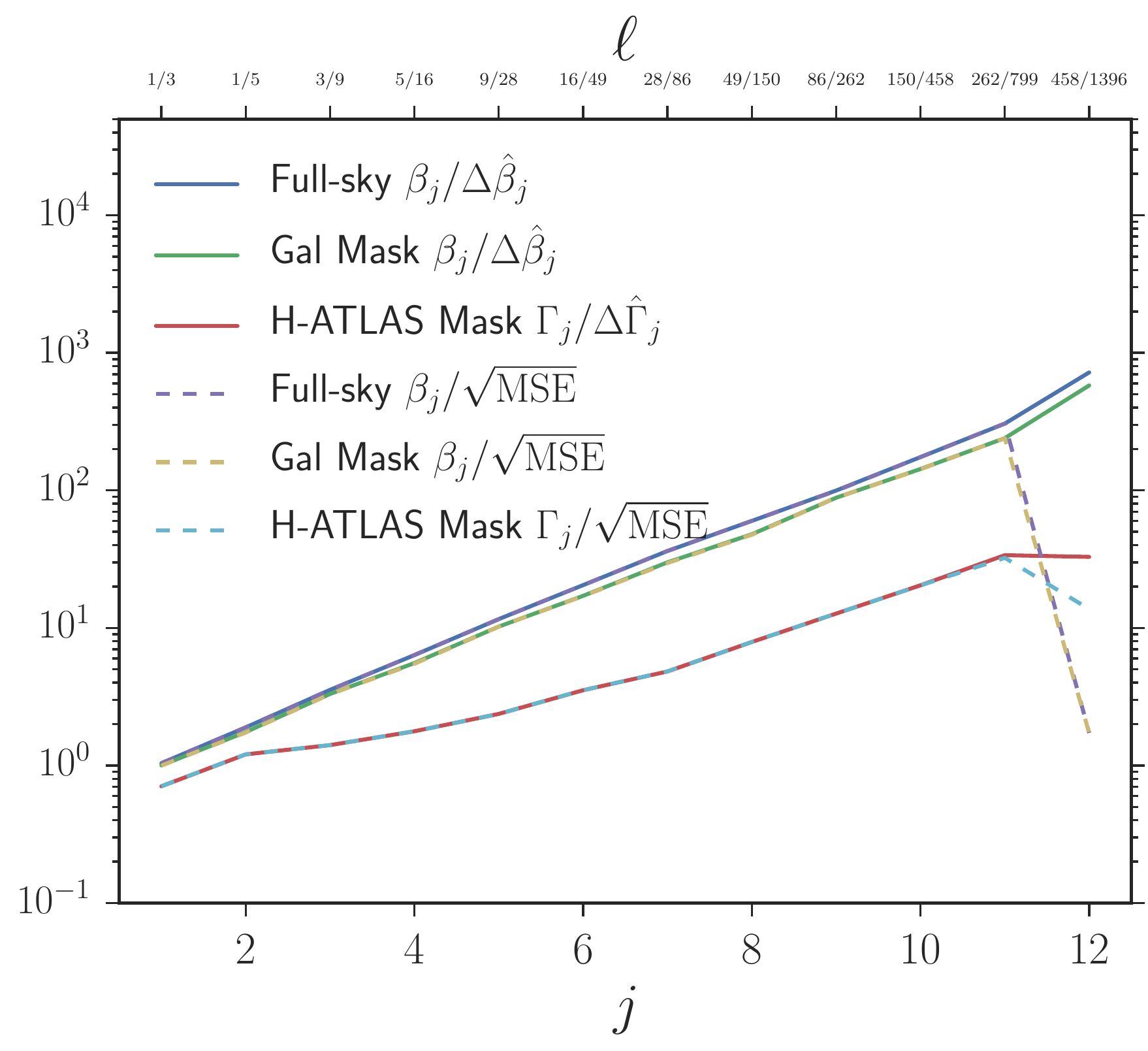}
  \includegraphics[width=0.5\textwidth]{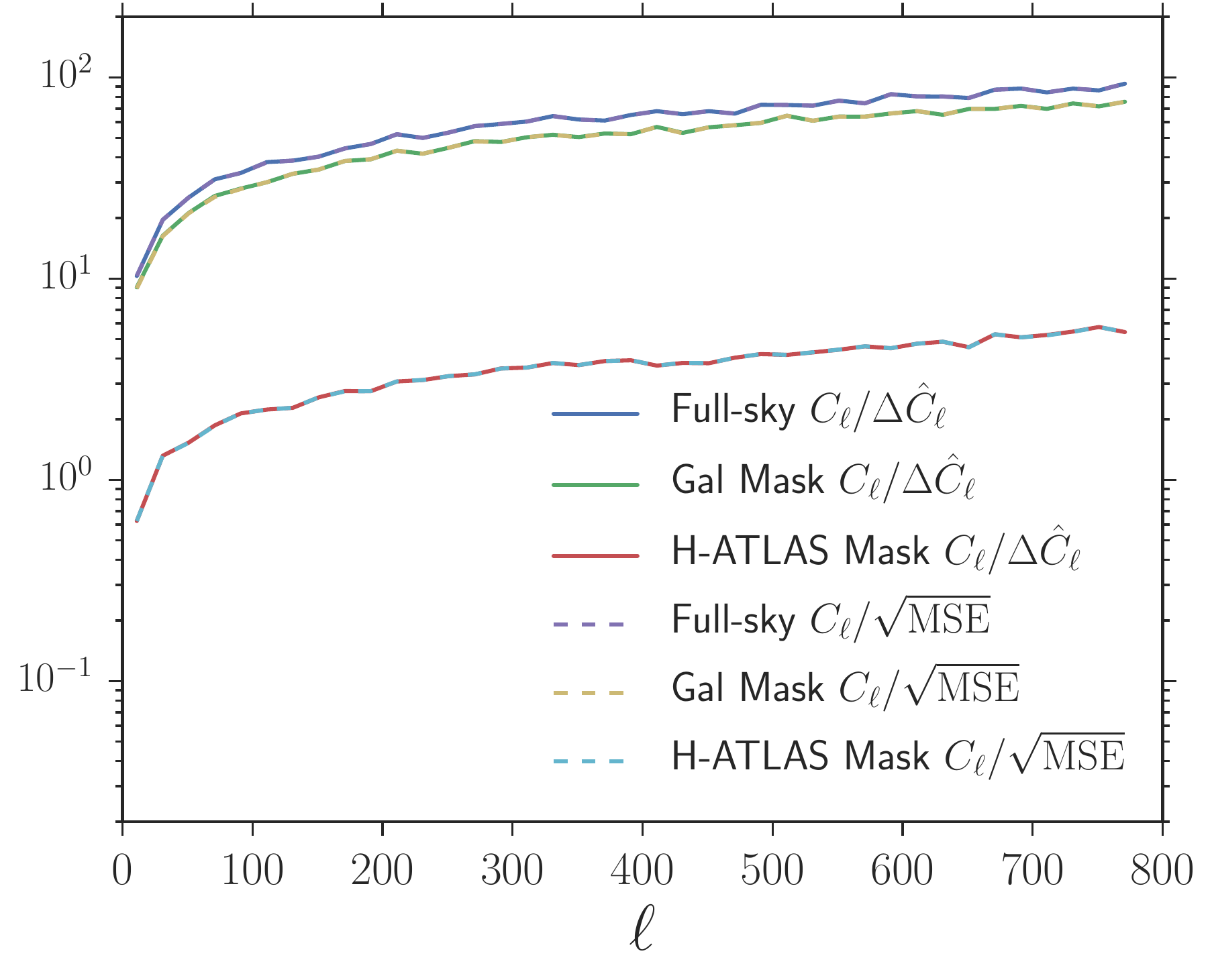}
\caption{\label{fig:metric} Different figures of merit discussed in text to assess the goodness of the estimators for the needlet (left panel) and harmonic (right panel) cases respectively. On top of the left plot
we quote the multipoles that roughly correspond to a given needlet frequency $j$.}
\end{figure}

The most important results are collected in figure~\ref{fig:metric} and Table~\ref{tab:S2N}, where we report the performances of the needlet- and harmonic-based methods, focusing on the MASTER corrections between the two approaches: to this end, we consider the the signal-to-noise ratio ($S/N$) as the relevant figure of merit. In particular, in figure~\ref{fig:metric} we present the $S/N$ \textit{per bandpower} for the two methods, evaluated as the ratio between the analytical expected value of the estimator (numerator) and a measure of variability, which can be either the standard deviation estimated from simulations, $\Delta\hat{\beta}_j \equiv \sqrt{\text{Cov}_{jj}}$, or the root mean square error (MSE), $\sqrt{\text{MSE}}$, where MSE $= \langle (\hat{\beta}_j - \beta_j^{\rm th})^2 \rangle_{\rm MC}$. The latter estimator takes into account also the possible presence of bias, but this is so small that the two measures are largely equivalent. Clearly, an higher value of this figure of merit entails a better performance of the estimator.

The performance of the needlet estimator seems to be rather satisfactory, with the figure of merit ranging from 1 to 3 for the H-ATLAS case and from 3 to 10 for the galactic case at the smallest frequencies $j=3,4,5$ (corresponding to multipoles of the order $\ell = 6, 10, 18$ respectively). At higher  frequencies, i.e. $j=10,11,12$ (corresponding to central multipoles of the order $\ell = 312,547,957$ respectively), the figure of merit is of order 200 (30) when the galactic (H-ATLAS) mask is applied. To make a rough comparison, the figures of merit for the standard power spectrum cross-correlation estimators are in the order of 9 at $\ell = 10$, and 80 at $\ell = 800$ for the galactic mask case, while the 
figure of merit in the H-ATLAS scenario goes from below 1 up to roughly 6 in the same $\ell$-range. 

To be fair, we stress that the numbers in figure~\ref{fig:metric} are not strictly comparable, because the bandwidths which are chosen for the standard harmonic domain estimator are constant 
%

across the multipoles domain and smaller than the equivalent needlet bandwidths, especially at high frequencies $j$. In addition, the $S/N$ computation sketched above does not include the impact of correlation among bandpowers that is quantified by off-diagonal elements in the covariance matrix. In order to overcome this issue we evaluate the \textit{total} $S/N$ of the cross-correlation detection by fitting the reconstructed needlet/harmonic power-spectrum for a free amplitude $A$ that rescales the theory template as $\hat{\beta}_j^{\kappa g} = A \beta_j^{\kappa g}$ (same for $C_{\ell}$'s) and estimating $S/N = \sqrt{\chi^2_{\rm null} - \chi^2_{\rm bf}}$. Here $\chi^2_{\rm null}$ is the chi-squared value of the fit under the null hypothesis (no cross-correlation), $\chi^2_{\rm bf}$ is the chi-squared values for the best-fitting model, and the full covariance matrix estimated with eq.~(\ref{eq:fullcov}) is used in the implementation. Total $S/N$ comparisons for the two methods are collected in Table~\ref{tab:S2N}.

Let us stress again that a direct comparison between the two approaches is far from being trivial due to their different coverage in the multipole space,\footnote{Recall that the highest multipole probed by imposing $j_{\rm max}=12 \,(11)$ corresponds roughly to $\ell \simeq 1400 \, (1160)$, while in the harmonic case the highest multipole is $\ell \simeq 780$.} but the least one can conclude from these results is that the two procedures have different advantages; in particular, we view as major assets for the needlet based algorithm the very high $S/N$ for aggressive masking and the natural choice of bandwidth parameters, while the advantage of the power spectrum based procedure seems the high resolution which can be achieved in multipole spaces.
\begin{table}[t]
\centering
\caption{Total $S/N$ comparison between the needlet- and harmonic-based methods for different observational setups. Note that both the H-ATLAS setups include MASTER corrections, while the numbers in parenthesis refer to total $S/N$ computed considering $j_{\rm max}=11$ for the needlet case.}
\begin{tabular}{ccc}
\toprule
\midrule
Setup & Harmonic & Needlet \\
\midrule
Full-sky & 9813 & 8768 (7830) \\
Galactic Mask & 8099 &  7163 (6376)\\
H-ATLAS Mask & 999 & 1042 (845)\\
\bottomrule
\end{tabular}
\label{tab:S2N}
\end{table}

\begin{figure}[H]
\centering 
\includegraphics[width=0.49\textwidth]{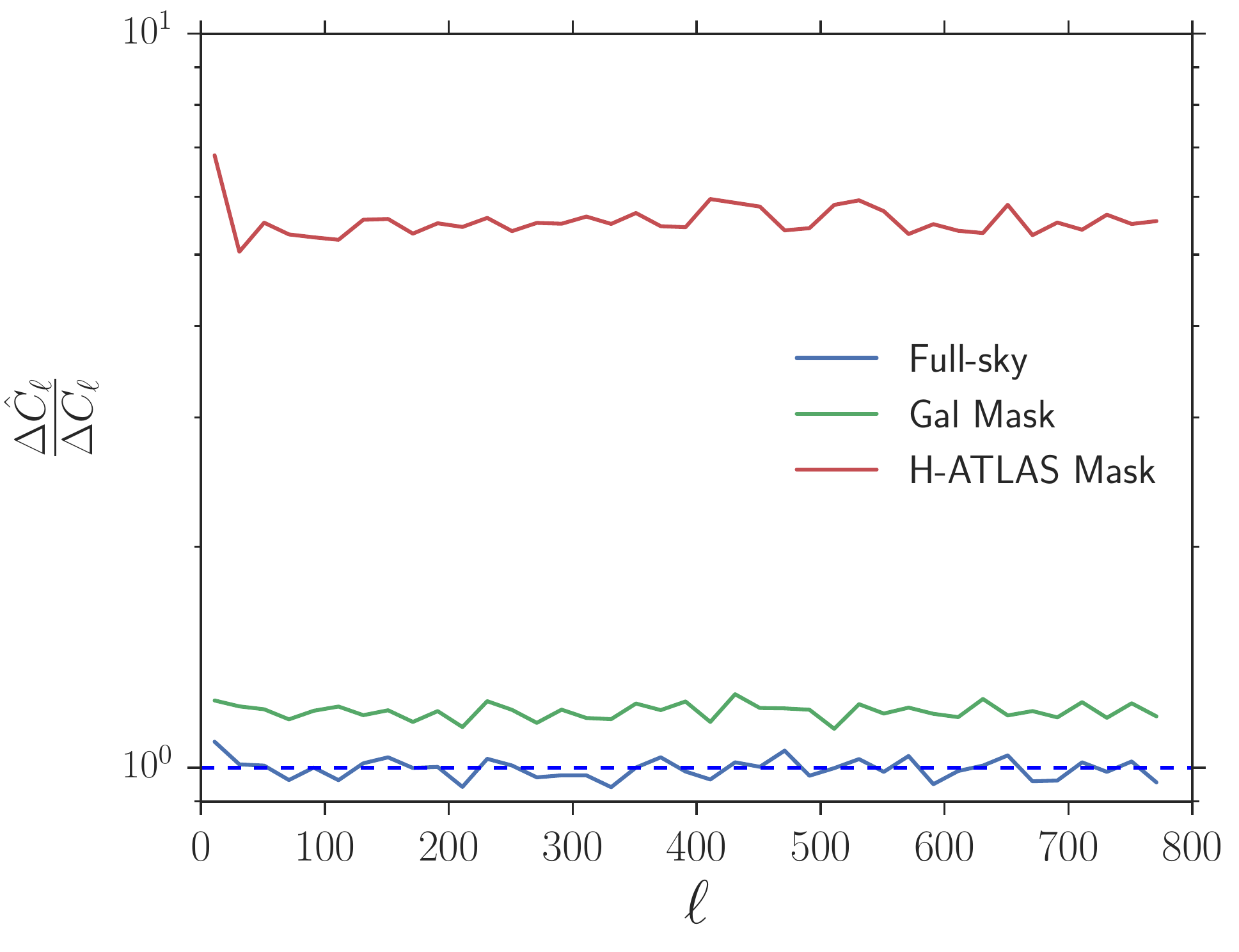}
\includegraphics[width=0.49\textwidth]{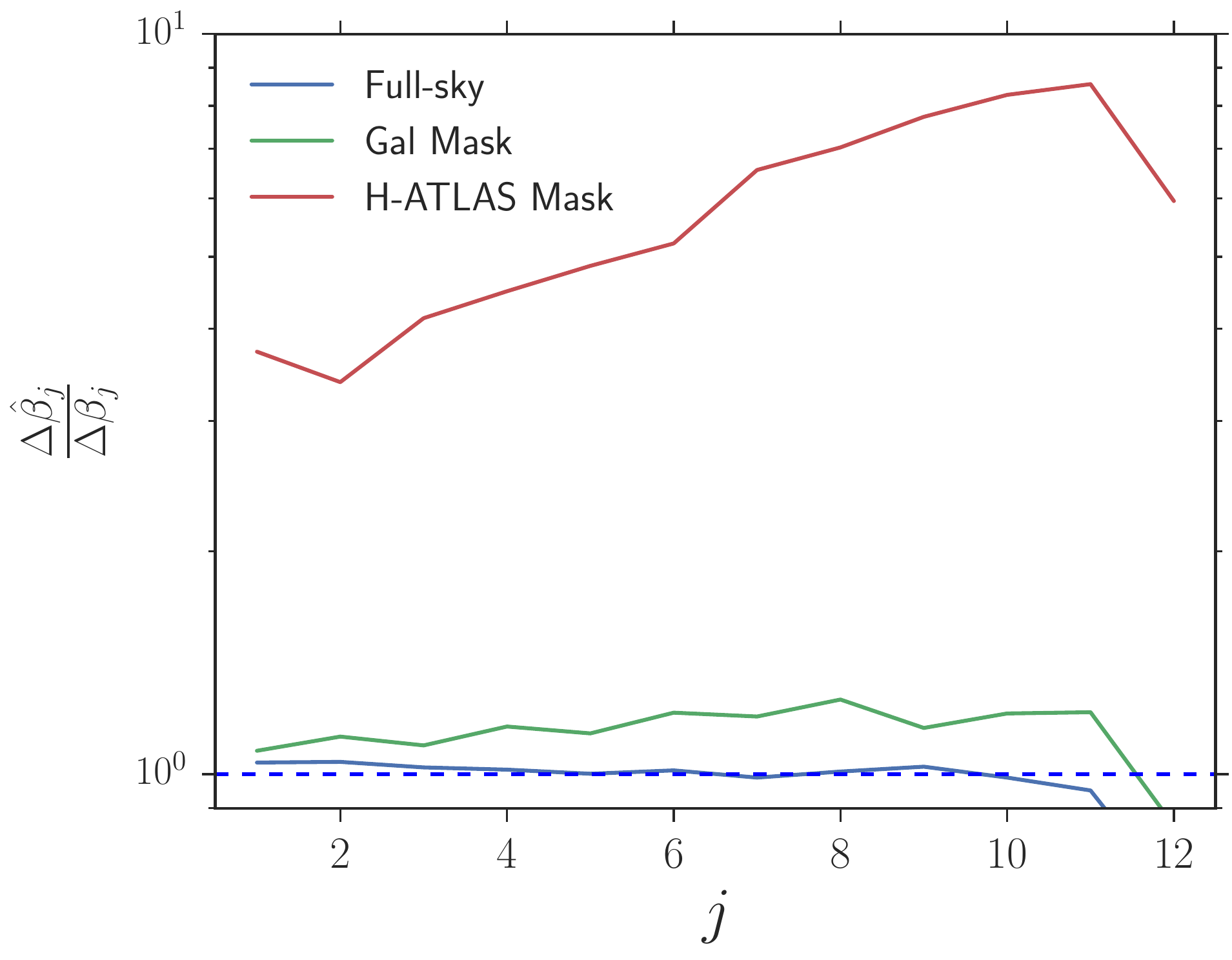}
\caption{\label{fig:var_ratio} The variance of cross-power spectrum estimates (left panel) and needlet 
cross-spectra (right panel) divided by the respective analytical full-sky variance. Note that we calculate the 
ratio  $\hat{\beta}_j/\Delta\beta_j$ for the H-ATLAS case (i.e. we consider the purple points in figure~\ref{fig:needsims}), not the pseudo spectrum $\hat{\Gamma}_j/\Delta
\Gamma_j$.}
\end{figure}

As a further check we show in figure~\ref{fig:var_ratio} the variance of the harmonic and needlet space 
estimators for the different observational setups as function of multipole $\ell$ and needlet frequency $j$,
normalized to the full-sky analytical variance. In order to be more quantitive on this aspect, we have investigated the relative scaling of the $S/N$ as function of the sky fraction between the two estimators and collected the results in Table~\ref{tab:S2N_fsky}. By comparing the expected $S/N$ in Table~\ref{tab:S2N_fsky} with the total ones measured as $\sqrt{\Delta\chi^2}$ and reported in Table~\ref{tab:S2N}, one can conclude that: (i) the rescaled $S/N$ are similar to the estimated ones for both the needlet and harmonic cases, suggesting the nearly optimality of the two methods; (ii)  the estimated $S/N$ are greater than the rescaled ones for both masks in the needlet case, while this is true just for the galactic mask in the harmonic approach; (iii) the needlet estimator has a better performance with respect to the PCL in the case of the H-ATLAS mask.

\begin{table}[t]
\centering
\caption{Relative scaling of the total $S/N$  between the two approaches for different observational setups. Note that both the H-ATLAS setups include MASTER corrections, while the numbers in parenthesis refer to total $S/N$ computed considering $j_{\rm max}=11$ for the needlet case, as in Table~\ref{tab:S2N}.}
\begin{tabular}{ccc}
\toprule
\midrule
Setup & Harmonic & Needlet \\
\midrule
$\left(\frac{S}{N}\right)_{\rm full-sky}\times \sqrt{f_{\rm sky}^{\rm Gal}}$ & 7961 & 7113 (6352) \\
$\left(\frac{S}{N}\right)_{\rm full-sky}\times \sqrt{f_{\rm sky}^{\rm H-ATLAS}}$ & 1123 & 1003 (896) \\
\bottomrule
\end{tabular}
\label{tab:S2N_fsky}
\end{table}
%
%

\section{Conclusions}
\label{sec:conclusion}
Cross-correlation analyses between independent cosmological datasets have the advantage to be 
potentially immune to any known (and unknown) systematics, as well as to extract signals hidden in noisy
data. In this way, cross-correlation measurements can provide us with a clearer view of the large scale 
distribution of matter, fundamental to reconstruct the dynamics and the spatial distribution of the 
gravitational potential that can be then translated into constraints on cosmological parameters, breaking
degeneracies with the astrophysical ones.

In this paper we begin a systematic analysis of the scientific potential associated to the expansion of the analysis domain in CMB-LSS cross-correlation studies to include the localization in the harmonic and spatial domains. In this initial application, by exploiting an ensemble of 
simulations, we have shown that under the same observational configurations the needlet spectral
estimator can enjoy some advantages over the harmonic one, thanks to the excellent needlets localization properties in both pixel and frequency space, as well as their optimal window function. 
Moreover, we have completed an initial needlet based analysis pipeline throughout the implementation of a novel MASTER-like approach for needlet spectral reconstruction in the case of aggressive masking 
($f_{\rm sky}\simeq 0.01$), reporting an higher total $S/N$ with respect to its harmonic counterpart. As we discussed earlier, these comparisons must be considered with some care, because the bin size is intrinsically different in the harmonic and needlet cases.

Motivated by these positive indications and results, in future research we plan to explore further the 
trade-off between $S/N$ and multipole 
localization, so as to achieve optimal bandwidth selection for a given experimental setting (such as the 
Euclid coverage mask). We also aim at applying this machinery to accurate CMB maps lensed 
with ray-tracing techniques \cite{Calabrese2015} and realistic galaxy mock catalogues based on N-body 
simulations by adopting, on the CMB side, the projected accuracy and sensitivity of forthcoming polarization oriented CMB probes, targeting the B-modes from cosmological gravitational waves and gravitational lensing. 
This work is of course preparatory for application to real data, from currently available LSS 
maps such as Herschel and WISExSCOS Photometric Redshift Catalogue (WISExSCOSPZ) 
\cite{Bilicki2016} to upcoming surveys such as Euclid, LSST, DESI, and WFIRST, in order to robustly 
extract cosmological information from cross-correlation measurements.

\acknowledgments
We are grateful to Carlo Baccigalupi and the anonymous referee for many useful comments on an earlier version of this paper.
A.R. and D.M. acknowledge support from ERC Grant 277742 Pascal. 
F.B. acknowledges partial support from the INFN-INDARK initiative.
Support was given by the Italian Space Agency through the ASI contracts Euclid-IC (I/031/10/0). 
In this paper we made use of \texttt{CAMB}, \texttt{HEALPix}, \texttt{healpy}, \texttt{matplotlib}, and \texttt{seaborn} packages.

\bibliographystyle{JHEP}
\bibliography{main}{}
\end{document}